\documentstyle[preprint,prc,aps,psfig]{revtex}
\newcommand{\be}{\begin{equation}}
\newcommand{\ee}{\end{equation}}
\newcommand{\bea}{\begin{eqnarray}}
\newcommand{\eea}{\end{eqnarray}}

\begin{document}
\renewcommand{\thefootnote}{\fnsymbol{footnote}}
\title{$J/\psi $--dissociation by a color electric flux tube}
\author{{\bf S. Loh, C. Greiner}\footnotemark[2]
{\bf and U. Mosel},\\[6mm]
$^{1}$ Institut f\"ur Theoretische Physik, Universit\"at Giessen,
\\
D-35392 Giessen, Germany.}
\footnotetext[2]{e-mail address: greiner@theorie.physik.uni-giessen.de}
\date{January 1997}
\maketitle
\thispagestyle{empty}
\renewcommand{\thefootnote}{\arabic{footnote}}
\begin{abstract}
We adress the question of how a $c-\bar{c}$--state (a $J/\psi $) can
be dissociated by the strong color electric fields
when moving through a color electric flux tube.
The color electric flux tube and the dissociation
of the heavy quarkonia state are both described within
the Friedberg-Lee color dielectric model.
We speculate on the importance of such an effect
with respect to the observed $J/\psi $--suppression in ultrarelativistic
heavy ion collisions.
\\[2mm]
\end{abstract}
\newpage
\pagenumbering{arabic}
\section{Introduction and Motivation}
Since the original work of Matsui and Satz
\cite{Mat86} who proposed $J/\psi $--suppression
as a clear signal for quark gluon plasma formation (the electric forces responsible
for the binding are sufficiently screened by the plasma so that no boundstate
should form) an intense work by several experimental groups
(NA38-collaboration) has adressed this
issue \cite{Exp}. Indeed a suppression in the corresponding dilepton signal
compared to the Drell-Yan background has been seen over the last years
for lighter projectiles like O+Cu, O+U and S+U. However,
these observations may also be explained alternatively:
(1) A part or even the total suppression can probably be explained
by $J/\psi $--absorption on the surrounding nucleons
\cite{Ger92},
i.e. $J/\psi + N \rightarrow \Lambda_C + \bar{D}$;
(2) additional absorption
by exothermic reactions like
i.e. $J/\psi + \rho \rightarrow D + \bar{D}$
might be attributed to `comovers' (`mesons')
being produced as secondaries \cite{Ga90}.
Indeed, if $\sigma _{abs}^{\psi N} \approx 7$ mb is assumed,
the reported
suppression could be nicely reproduced by the absorption within the nuclear
environment.
In any case, the general consensus has then been that a highly (energy) dense
intermediate reaction zone was needed to explain the observed suppression.

Recently the data taken in Pb+Pb reactions at 160 AGeV
by the NA50-collaboration \cite{Exp1} have given new excitement
to the whole issue as a stronger absorption has been found than suggested
by the models of absorption models on nucleons.
It was immediately speculated \cite{Bl96}
that a large inner region of the reaction zone must absorb {\em all}
$J/\psi $ (being `black') in order to be compatible with observation.
Though it was reported by Gavin and Vogt \cite{Ga96}
and by Cassing and Ko \cite{CK96} that the comover model also
could explain the stronger suppression as relatively more mesons are
produced for large $E_{t}$ events (or fluctuations) than for lighter systems,
the latter calculations also show that this suppression takes place
at energy densities that are much higher than those in which normal
hadrons are expected to survive.

In popular microscopic models, simulating the whole reaction
of the collision, in the first few moments interaction
strings are produced which subsequently decay (fragment) into secondaries
(mesons, baryons, string-like hadronic resonances). These are best visualized
as color electric flux tubes. Their creation is thought to happen by the exchange of a
color octet gluon or by colorless momentum transfer
among target and projectile nucleons. (The color sources in the endcaps
should be thought as a quark on the one side and an anti-quark on the other side,
or as a quark and a di-quark, respectively.) These strings might also overlap
(if their density is large) to form higher charged tubes,
so-called color ropes \cite{Bi84}.
In any case, before a further hadronic-like or partonic-like state of matter does form, such
a temporary build up and decay of strings is
assumed to describe the
very early collision phase. In fact one might call such an
environment
a precursor of quark matter.

The $J/\psi $ (or better $c\bar{c}$-state which later will form the $J/\psi $)
as a rather heavy hadronic particle is also produced at the earliest state of the reaction
in a hard collision among the nucleons. Thus it is natural to ask what happens if
a (pre-) $J/\psi $--state moves (or enters) into an environment of color electric strings.
As the string carries a lot of internal energy (to produce the later secondaries)
the quarkonia state might get absorbed and completely dissociated by the
intense color electric field inside a single flux tube. This intuitive idea
is the basic reason for our work.

The dynamics of the flux tube we describe within the Friedberg-Lee
model \cite{FL77,Wilbuch}
by a semiclassical transport algorithm (section II). As a first step, within this model,
we generate a bound $c\bar{c}$--state with reasonable properties
in order to mimic a $J/\psi $ (section III). We then adress how such a state gets
dissociated when inserted into a chromoelectric flux tube
and how the tube subsequently does split (section IV).
We conclude with a summary and the implications of our investigation
on the issue of $J/\psi $--suppression.
\section{Dynamical realization of the Friedberg-Lee model}
Recently we have written down semiclassical transport equations for quark distributions
\cite{Vet95,Lo96} starting
from the phenomenological Lagrangian of Friedberg and Lee \cite{FL77,Wilbuch}
\be
  \label{lagrange}
  {\cal{L}} = \bar{\Psi}(i\gamma_{\mu}\partial^{\mu}-m_0-g_0\sigma)\Psi
              + {1 \over 2}(\partial_{\mu}\sigma)^2 - U(\sigma) 
              - {1 \over 4} \kappa(\sigma)F_{\mu\nu}^aF^{\mu\nu}_a 
         - ig_v\bar{\Psi}\gamma_{\mu}{\lambda^a \over 2}\Psi A^{\mu}_a \; ,
\ee
where $\Psi$ denotes the quark fields, $\sigma$ is the color singlet scalar
field representing the long range and nonabelian effects (of multi gluon
exchange), and the last term contains the interaction of the residual
classical and abelian color fields $A_{\mu}^a$.
All the nonabelian effects are assumed to be absorbed in the color
dielectric function $\kappa(\sigma)$ which is chosen such that
$\kappa$ vanishes as $\sigma$ approaches its vacuum value $\sigma = \sigma_v$
outside the bag and $\kappa = 1$ inside,
\be
  \kappa(\sigma) = |1 - {\sigma \over \sigma_{vac}}|^2
                   \Theta(\sigma_{vac} - \sigma) \; .
\ee
\be
 U(\sigma) = {a \over 2!}\sigma^2 + {b \over 3!}\sigma^3 +     
             {c \over 4!}\sigma^4 + B    
\ee
is the self-interaction potential for the scalar
$\sigma$-field containing cubic and quartic terms.
Details of the numerical realization and the further approximations made are found
in \cite{Lo96,Lo96a}. One obtains two transport equations for the
phase space distributions of quarks ($f$) and antiquarks ($\bar{f}$)
\bea
 & & (p_{\mu}\partial^{\mu}-m^*(\partial_{\mu}m^*)\partial^{\mu}_p)f(x,p) =
  g_vp_{\mu}F^{\mu\nu}\partial_{\nu}^pf(x,p)  \nonumber \\
  \label{vlas2}
 & & (p_{\mu}\partial^{\mu}-m^*(\partial_{\mu}m^*)\partial^{\mu}_p)\bar{f}(x,p) =
  -g_vp_{\mu}F^{\mu\nu}\partial_{\nu}^p\bar{f}(x,p) \; ,
\eea
which are numerically solved by applying the test-paticle method,
a classical equation of motion for the scalar soliton field
\be
  \label{klegor}
  \partial_{\mu}\partial^{\mu}\sigma + U'(\sigma)
  + \frac{1}{4} \kappa'(\sigma)
  F_{\mu\nu}^cF^{\mu\nu}_c + g_0\bar{\Psi}\Psi = 0
  \;,
\ee
which is solved using a
staggered leapfrog algorithm
\cite{Vet95},
and an equation of motion for the confined colourelectric field
$\vec{D}=\kappa \vec{E}$
\be
  \label{maxsca}
  \vec{\nabla}(\kappa\vec{\nabla}A_0) = -j_0 \; ,
\ee
which is determined by a
two-dimensional finite element method
(appropriate for cylinder symmetrical configurations) developed by Mitchell
\cite{Mi88}.

First we note that here we explicitly suppress the color magnetic fields, i.e.
$F_{\mu\nu}^cF^{\mu\nu}_c \rightarrow 2 \vec{E}^2$.
(In a string like configuration with nearly constant radius Wilets and Puff
have shown that due to the displacement current in the Maxwell equation magnetic
fields are not produced \cite{Wi95}.) $f$ and $\bar{f}$ describe the distribution
of charged quarks and anticharged antiquarks (or diquarks) so that the colour
charge density is expressed by
\be
  \label{coldens}
   j_0 = {\eta \over (2\pi)^3} \int d^3p (f(x,p)-\bar{f}(x,p)) \; ,
\ee
whereas the scalar density $\rho_s \equiv \bar{\Psi } \Psi $ entering into
eq. (\ref{klegor}) reads
\be
  \label{sdens}
  \rho_s = {\eta \over (2\pi)^3} \int d^3p {m^* \over \omega}
           (f(x,p)+\bar{f}(x,p)) \; .
\ee
($\eta \equiv 4$ accounts for spin and flavour degeneracy.)

Within this approach one succeeds in simulating dynamically confinement
as
(a) a single charged quark is forced back
into the configuration by the dielectric displacement $\vec{D}$
if it tries to escape \cite{Wilbuch}; and (b) the dielectric displacement
$\vec{D}=\kappa \vec{E}$ stays also within the configuration as it has to
vanish in the outside (and nonperturbative) vacuum where
$\kappa \rightarrow 0$ \cite{tdlee}.

Static properties of soliton-like nucleons and mesons are succesfully
reproduced by an appropriate fixing of the scalar potential $U(\sigma )$
and the coupling constant $g_s$. The vector coupling
$\alpha _s \equiv g_v^2/4\pi = 1.92$ is fixed
by assuming an effective string constant $\tau = 1$ GeV/fm
when simulating a colourelectric flux tube \cite{Lo96}.
Here the volume averaged electric field
$\langle E_z \rangle $ takes a value of 1.6 fm$^{-2}$
with $E_{z,max} =2.2$ fm$^{-2}$ along the z-axis. The radius
of the string results as approximately 0.8 fm.
\section{$J/\psi $- and $D$,$\bar{D}$-states}
As a first step, within our model, we want to describe semiclassically
a bound $c\bar{c}$-state (synonymously a $J/\psi $) and also
a bound $c\bar{q}$-state (synonymously a $D$).
We start from a light meson state described within our model
(using the parameters of the model
\cite{Vet95,Lo96,Lo96a}) and then adiabatically
increase the mass of the initial light quark(s)
($m_{i,q} \approx 10$ MeV) to a final heavy quark mass $`m_{f,c}'=1.56$ GeV
(taken from ref. \cite{Mat86}), i.e.
\be
\label{timem}
\begin{array}{rll}
m_q(t) = & m_{i,q}(t=0) + (m_{f,c}-m_{i,q}(t=0)) t/\tau _{s.o.}
& \, \mbox{for} \, \, 0<t \leq \tau _{s.o.}  \\
m_q(t) = & m_{f,c}
& \, \mbox{for} \, \, t > \tau _{s.o.}   \nonumber
\, \, \, .
\end{array}
\ee
The change of the meson state with mass of both (or one) quarks is thus
described fully dynamically.
In figs. 1 we see how the bag size
shrinks from $r_{cms} \approx 0.66$ fm to $r_{cms} \approx 0.25$ fm
for the case of a $c\bar{c}$-groundstate
and to $r_{cms} \approx 0.40$ fm for the case of a $c\bar{q}$-groundstate
for a switching on time of $\tau _{s.o.}=20$ fm/c.
We checked our results
for various switching on times
$\tau _{s.o.}=20,30,50$ fm/c and found
that the state described does
not depend on the parameter $\tau _{s.o.}$.
The shrinking of the states is in
accordance with the expectation of any baglike model. When the quark masses
are increased to their final value, the pressure the more heavy quarks assert
on the surface drops too, thus leading to the shrinking of the whole bag
until it stabilizes at a smaller radius where pressure equilibrium
is reached again.
The radius of the heavy meson
states (the `$J/\psi $'- and `$D$'-meson) are in the right range of what
one would expect from other quarkonia model.
In addition, fig. 2  shows the spatial quark distribution and the soliton solution
$\sigma (r)$ as a function of the radius for the $c\bar{c}$-state
at the beginning and the end of the evolution according to (\ref{timem}).

When determining the overall
mass of the states we find for $m_c=1.56$ GeV
\bea
   \label{mass}
     m_{c\bar{c}} & = & 4.10 \, \mbox{GeV} \\
     m_{c\bar{q}} & = & 2.25 \, \mbox{GeV} \, \, \, .\nonumber
\eea
These numbers have to be compared to (a) the mass of the
$\eta _c$: $m_{\eta_c}=2.94$ GeV and the $J/\psi $: $m_{J/\psi }=3.097$ GeV
and (2) to the mass of the $D$-meson: $m_D=1.89 $ GeV and the
$D^*$: $m_{D^*}=2.01$ GeV. The $c\bar{c}$-state generated
within the model (and within the parameters chosen to describe the light mesons)
are thus about 1 GeV too heavy, the $c\bar{q}$-state about 0.4 GeV, whereas
the difference $2m_{c\bar{q}}-m_{c\bar{c}} \approx 0.4 $ GeV is in
quite good agreement to $2m_D- m_{J/\psi }=0.68 $ GeV.
This discrepancy one can relax by choosing a smaller
bare mass of the charm quark. This mass is rigorously not known and only
extracted by various charmonium models and might vary between
1 to 1.6 GeV \cite{PD94}.
Taking the a priori unknown charm quark mass as $\approx \, 1.25$ GeV
(compare figs. 1),
the mass of the $c\bar{c}$-state would lie slightly above the experimental
value by about $\approx 200$ MeV, whereas the mass of the $c\bar{q}$-state
is lowered by about the same value, thus providing a much better
description within our model.
In the following we stay within our model states (\ref{mass}) (described
by a bare mass of $m_c=1.56$ GeV) as our conclusions will not depend
on their masses and the results given in the next section
will only weakly be affected
by assuming a different mass for $m_c$ within the afore mentioned range.
\section{$J/\psi$-dissociation by a color flux tube}
We now turn to the main issue of the present work:
What happens to a $c\bar{c}$-state entering the region spanned by
a single chromoelectric flux tube?

We prepare a chromodielectric flux tube
(see also \cite{Lo96}) by pulling the
opposite charge distributions of a (light) groundstate meson
steadily apart. The quarks are propagated with
a constant velocity up to a final spatial extent of about 8 fm (see e.g.
fig. 3) within a time interval of about 13 fm/c. At that given time
$t_0 \approx 13$ fm/c we keep the quark distributions at the endcaps {\em fixed}, thus
providing the sources for the chromodielectric field.
We then insert the `$J/\psi $' `by hand' into
the central interior of the (by now) stationary flux tube,
where the initial momentum and spatial distributions of the heavy quarks
are chosen as the one obtained in the previous section. Initializing
the cylindrically symmetric configuration in this way, we proceed by solving
the full set of dynamical equations of motion, i.e. the two
Vlasov equations (\ref{vlas2}) for the heavy quark and antiquark distribution,
the equation (\ref{klegor}) for the
soliton field $\sigma $ and the one for the electric potential (\ref{maxsca}).

In fig. 3 we show the evolution of the total electric field energy
as a function of time. As one notices, after inserting the bound state,
the field energy decreases approximately linearly with time. What happens is that the
heavy quark (described by its phase space distribution) is steadily
pulled to one endcap of the string carrying opposite charge,
whereas the antiquark carrying the opposite charge is correspondingly
pulled into the other direction (see figs. 4). Being pulled apart, the two
displaced charges screen the overall electric field in between.
In other words, the electric field energy originally stored
is transformed into the kinetic energy of the oppositely moving
two heavy quarks. Being accelerated, after some smaller time interval
$\Delta t \stackrel{<}{\approx }1$ fm/c, both
heavy quarks have nearly reached the speed of light, so that the linear
decrease of the overall electric field energy depicted in fig. 3 becomes
obvious.

In figs. 4 also the evolution of the $\sigma $-field is depicted
for various time stages. At the initial times one clearly sees how
the soliton solution of the flux tube is distorted by introducing
the $c\bar{c}$-state into the center of the string.
At the timesteps up to about 18 fm/c the $\sigma $-field follows the
quark distribution quite well, its shape reflects the shape of the
latter distribution. At 20.8 fm/c the inner heavy charges have reached
the fixed endpoints of the flux tube. At this point we stop
-- for numerical reasons --
the further propagation of the quarks and concentrate on the further
time development of the $\sigma $-field shown in figs. 5.
In this
further evolution the soliton field relaxes much more slowly to its
vacuum value compared to the more or less instantanous screening
of the electric field. It typically takes about 5 fm/c until $\sigma $
turns for the first time to its vacuum value \cite{Lo96a}.
This long time interval
is basically a consequence of the nonlinear potential $U(\sigma )$
being flat around $\sigma \approx 0$ and corresponds to the time
needed for the soliton field to `move down' the potential hill to
its vacuum value at $\sigma = \sigma_V$. Even when $\sigma_V$ has been reached, though,
the field still starts to oscillate around this value which can be seen
in figs. 5: In the second row, at t=23.2 fm/c, the $\sigma $-field value is
slightly above $\sigma _{vac}$, whereas in the third row, at t=26.8 fm/c, its
value is nearly in the perturbative region.
The original field
energy carried by the soliton field (the `Bag' energy) being still present
after the heavy quarks have already moved apart is initially more or less
conserved. The late oscillations observed in the simulations are
only damped by a transverse expansion of the field itself, which,
however, is a rather slow process because of the large `glueball' mass
of the field.
Thus it takes up to $\approx $ 15 fm/c until the $\sigma $-field has finally
relaxed and the static heavy $D/\bar{D}$-meson states are formed.
If one would consider a chiral O(4)- extension
for the soliton field one expects that these oscillations
might accordingly transform into low-momentum (and nearly massless)
pion modes and should accordingly be damped away more quickly.

From our simulation we conclude that
a localized $c\bar{c}$-state immediately gets separated by the strong
colourelectric field, and, in return, will finally be dissociated
into $D$-meson (or $\Lambda _C$-baryon) like configurations.
One wonders if such a dissociation would persist also if initially
the $c\bar{c}$-state enters the string with some moderate transverse momentum.
The finite element method implemented in order
to calculate the chromoelectric potential $A_0(r_t,z)$ can presently
only handle cylindrically symmetric configurations (as e.g. a static string).
Hence we are yet not able
to simulate the full dynamics when a $c\bar{c}$-state (as generated
in the previous section) enters a static and elementary flux tube with some transverse
momenta and finite impact (though ths work is in preparation \cite{Tr96}).
In the following we give an estimate that such a disintegration
of a $c\bar{c}$-state should still persist for even quite large transverse
momenta.

The time needed to pass the whole flux tube is roughly estimated as
\be
   \label{estm}
   \Delta t \, \approx \, \frac{2 r_{cms}}{v_{t}^{J/\psi }}
  \, \approx \, 2 r_{cms}\frac{\sqrt{p_t^2+m_{J/\psi }^2}}{|p_t| }
  \, \approx \, 2 \, - \, 4 \, \mbox{fm/c}
\ee
for moderate to large (transverse) momenta in the range from 1 to 5 GeV/c.
For an average field strength $\langle |E_z|\rangle$ of 1.61/fm$^2$ the heavy quarks
will suffer a gain in longitudinal momentum given approximately by
($\alpha _s =g_v^2/4\pi\approx 2$)
\be
  \label{mom}
  |\Delta p_z| \approx g_v|E_z| \Delta t \approx 1.6 \,
  \frac{\mbox{GeV}}{\mbox{fm}} \Delta t \approx 3\, -\, 6 \,\mbox{GeV/c}
  \, \, \, ,
\ee
whereas its internal rise in longitudinal kinetic energy of both
quarks (by assuming that each heavy quark moves,
on the average, with, let us say, half speed of light)
\be
  \label{energy}
  \Delta W \approx 2g_v|E_z| \Delta s
  \approx g_v|E_z| \Delta t
  \approx 3\, -\, 6 \,\mbox{GeV}
  \, \, \, .
\ee
Such an increase in momenta or internal kinetic energy is certainly
sufficient to break up the $c\bar{c}$-state into two D-meson like
states, as the energy needed is given by the difference in mass,
i.e. of about 0.68 GeV. As the strings are thought to be energetic enough
to produce all secondaries over the whole rapidity range in the
heavy ion collision, its internal color electric field
should be large enough to break up even more heavily bound systems
like the $J/\psi $. We are hence quite confident that our
observed dissociation would also be found within a full 3-dimensional
simulation of a $c\bar{c}$-state interpenetrating a chromodielectric
flux tube with a nonvanishing transverse momentum.

In some popular charmonium models a rather strong effective Coulomb
potential between the
$c$ and $\bar{c}$ quark of the form
$V_{c\bar{c}} \,  \sim \, -C/r$
, to be valid at short distances, gives a dominant contribution to
the overall spectra and gives rise
to a binding energy estimated to be 800 MeV for the $J/\psi$
as a 1s-state. When modeling this state within our semiclassical approach, the
groundstate is only bound by the strong confining scalar field.
The dissociation described above corresponds to classical field ionization.
The same effects exist in a quantum mechanical treatment:
A quantum
mechanical system of two heavy quarks will dissociate
at long times if a
constant background field with a potential
$V\sim -g_vEz$ is switched on \cite{Opp28}.
For our specific numbers, along the
z-direction, this leads to a potential of
$V \approx (-1.7 )z\, \mbox{GeV/fm} \, - \, (-2.1)z \, \mbox{GeV/fm}$
due to the strong electric field.
The modifications are thus {\em strong}
and {\em nonperturbative}. For $z=0.3$ fm the
effective total potential
$V_{eff}(z) \sim - C/z -g_v Ez$
is already lowered by
$\approx $ 600 MeV compared to the `bare' potential $-C/z$.
The existence of a (then highly modified) resonant 1s-state seems
very unlikely.
In addition one also can consider
this quantum mechanical problem as a nonstationary situation when the $J/\psi$
passes through the tube with some transverse velocity. When entering, the abrupt change in situation
can not be considered as adiabatic, but better as diabatic:
The
original bound state has no time to adiabatically lower
its energy by at least 0.6 GeV, but will immediately dissolve into
the continuum.

\section{Conclusion and Outlook}
In this work we have adressed the question of how a $c\bar{c}$-state
(a $J/\psi $) behaves inside a chromoelectric flux tube.
The flux tube as well as the $c\bar{c}$-groundstate are described within
the model by Friedberg and Lee \cite{FL77}.
Due to the strong color electric field
inside the flux tube the heavy meson becomes dissociated rather immediately
(on a timescale of $\stackrel{<}{\approx }1$ fm/c) ending up finally in
$D$- or $\Lambda_c$-like states. We could not simulate a fully 3-dimensional
situation when the heavy meson enters the string with some fixed transverse
momentum because our present numerical realization of the model works
only for cylindrically symmetric configurations. However, we give an
estimate that such a dissociation should also occur
by field ionization for moderate and higher transverse
momenta of the $c\bar{c}$-state when passing through the tube within our
semiclassical approach.

We believe that this investigation adresses an intriguing question
for understanding particular topics of QCD with respect to
relativistic heavy ion collisions. If one believes in the succesful descriptions
of microscopic approaches
(like FRITIOF \cite{An87}, RQMD \cite{So89}, VENUS \cite{We89}, HSD
\cite{Ca96}), a large
region in space in the first few moments after the reaction is spanned
by highly excited longitudinal strings. These strings will later materialize into
all the secondaries and the total energy carried by them, thus
leading to a more or less complete distribution of the energy between
target and projectile rapidity. In this sense the flux tubes should be interpreted as
highly energetic excitations of the QCD vacuum. One might consider such
regions as a precursor of a QGP and deconfinement. Especially for the more
heavy systems the effective region (or volume) of all the strings
being produced within a short time interval of less then one fm/c
gets so large that the strings become closely packed or already
overlap. Although most often the strings are thought to fragment
independently one might also consider the possibility
of color rope formation as higher charged tubes.
For a quantitative estimate we depict in fig. 6
the position of nucleonic strings (highly energetic strings excited by
a target and projectile nucleon)
all being produced in a cms-time interval
of $\approx 0.5 $ fm/c within a slab of the complete transverse area
and 1 fm thickness along the longitudinal direction
in the cms frame in a central S+U reaction at 200AGeV generated
within the HSD approach \cite{Ca96}. In total about 58 strings are produced
in the transverse area where the sulphur ion faces the uranium nucleus.
This area is about 40 fm$^2$. If one assumes that the strings are homogeneously
distributed (which they are not) this corresponds to a mean area of
0.7 fm$^2$/string. Hence all strings are closely packed if the radius
of a string is about 0.5 fm (in the Friedberg-Lee model it turns
out to be nearly 0.8 fm). This simple reasoning illustrates that
the whole reaction area in an ultrarelativistic heavy ion collisions
might be completely filled by strings in the first few moments.
If such a large spatial region does exist, a large
fraction of the produced $J/\psi $'s have to pass it and thus can
be affected by this highly excited environment.
This will depend crucially on the (average) length of the produced strings before
they hadronize, as this decides about the available `empty' space
for the $J/\psi$'s (or pre-$J/\psi$-states) to escape the initial reaction zone
without entering any of the individual strings being built up.
If they will get
dissociated (in part or all of them) this would lead to a
minor or stronger suppression of $J/\psi $ to be finally observed.
Such a conclusion can only be further tested and strengthened
if one incorporates these ideas within one of the present
microscopic transport algorithms.
\\[15mm]
{\bf Acknowledgements:}
The authors would like to thank W. Cassing for providing
us with fig. 6, for stimulating discussions and a careful reading of
the manuscript.
We acknowledge support by the BMBF and GSI Darmstadt.
\\[10mm]
\parskip0mm
\par
\newpage
\parindent 0mm
\begin{figure}
\vspace{0cm}
\centerline{\psfig{figure=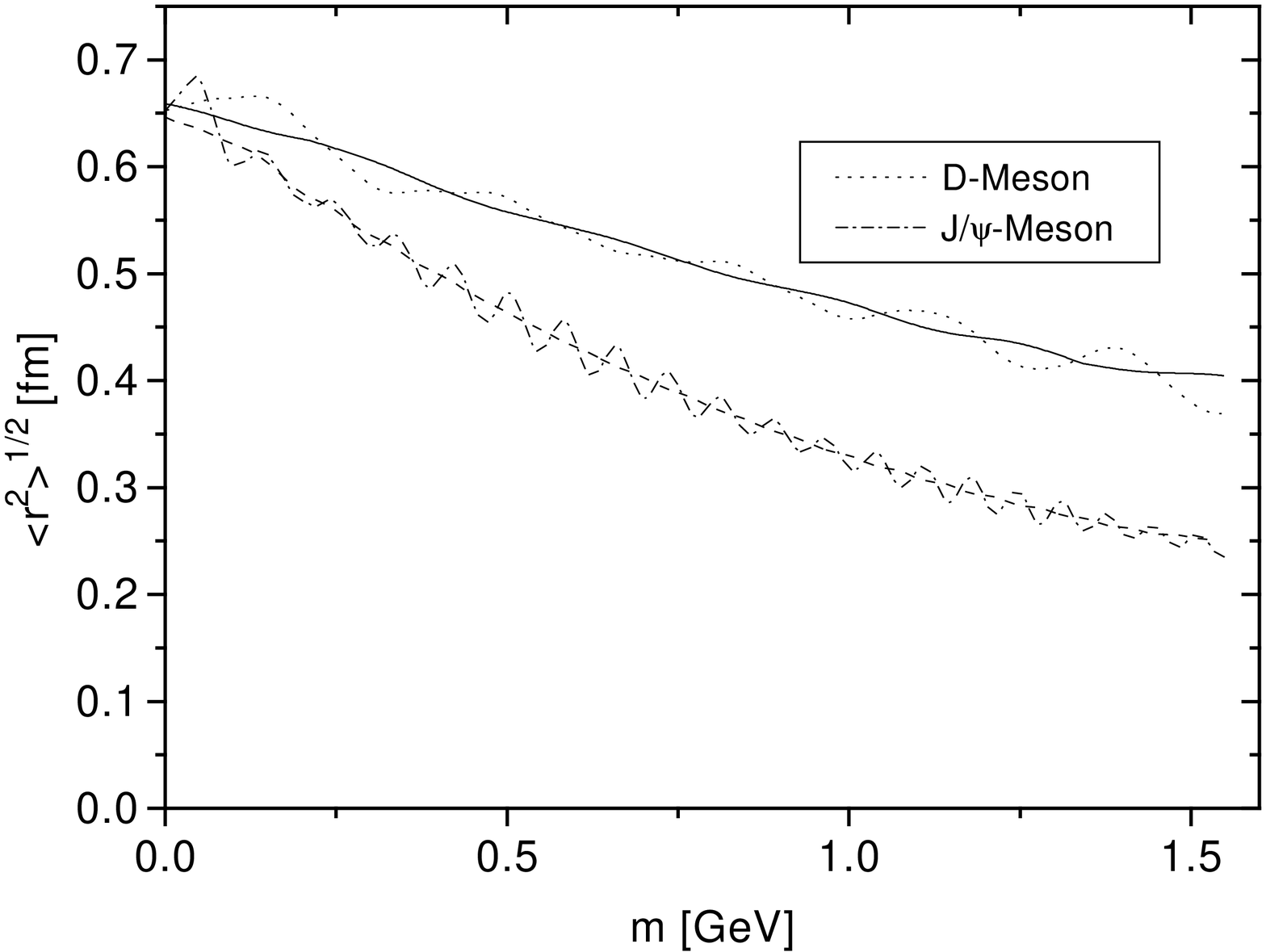,width=12cm}}
\vspace{0cm}
\end{figure}
%
%
\begin{figure}
\vspace{0cm}
\centerline{\psfig{figure=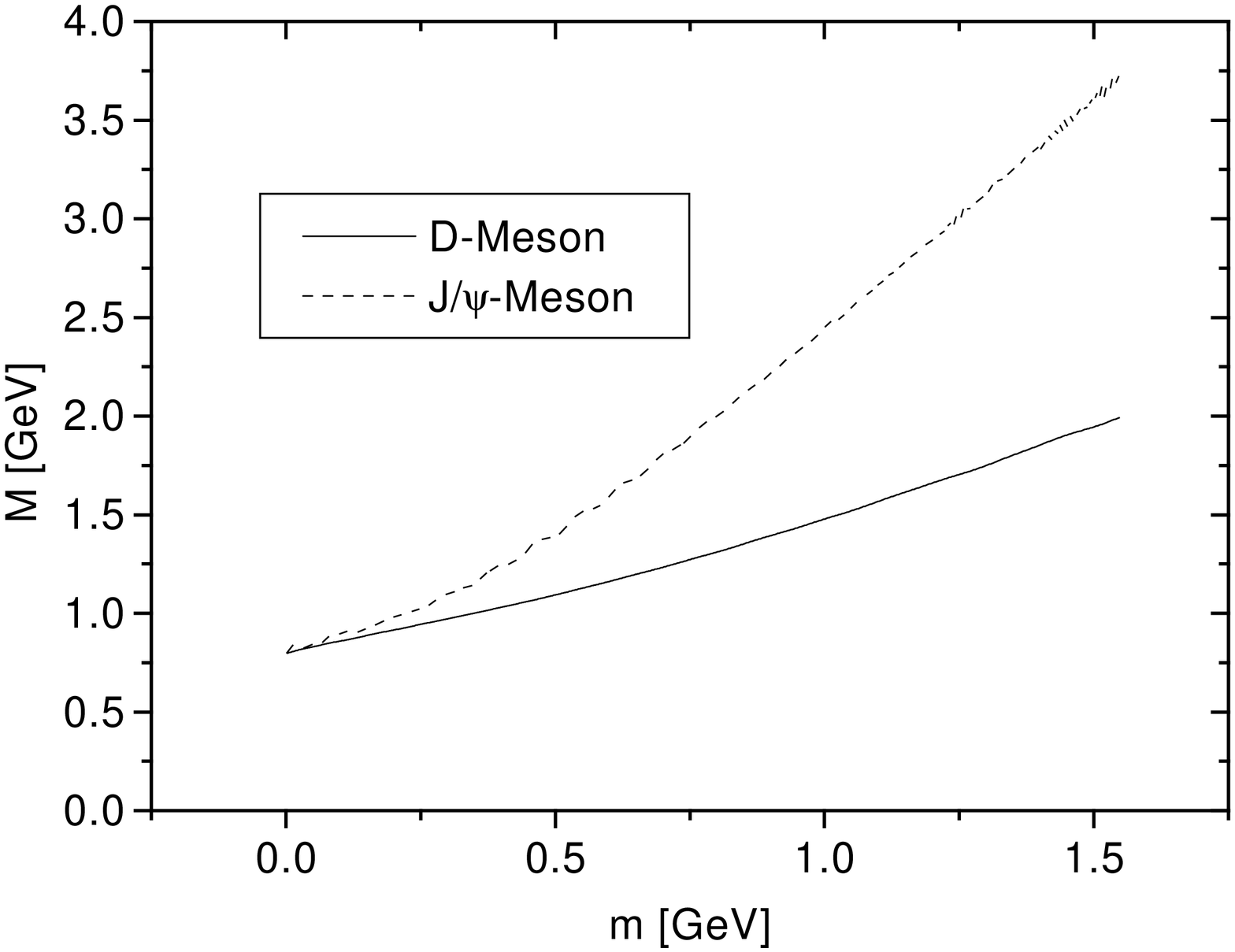,width=12cm}}
\vspace{1cm}
\caption{The average radius $\langle r^2 \rangle^{1/2}$ and the
overall mass M of the
the $c\bar{c}$-state and the
the $c\bar{q}$-state are shown
as a function of the bare charm quark mass $m_q$.}
\label{figure1}
\end{figure}
\newpage
\newpage
\begin{figure}
\vspace{0cm}
\centerline{\psfig{figure=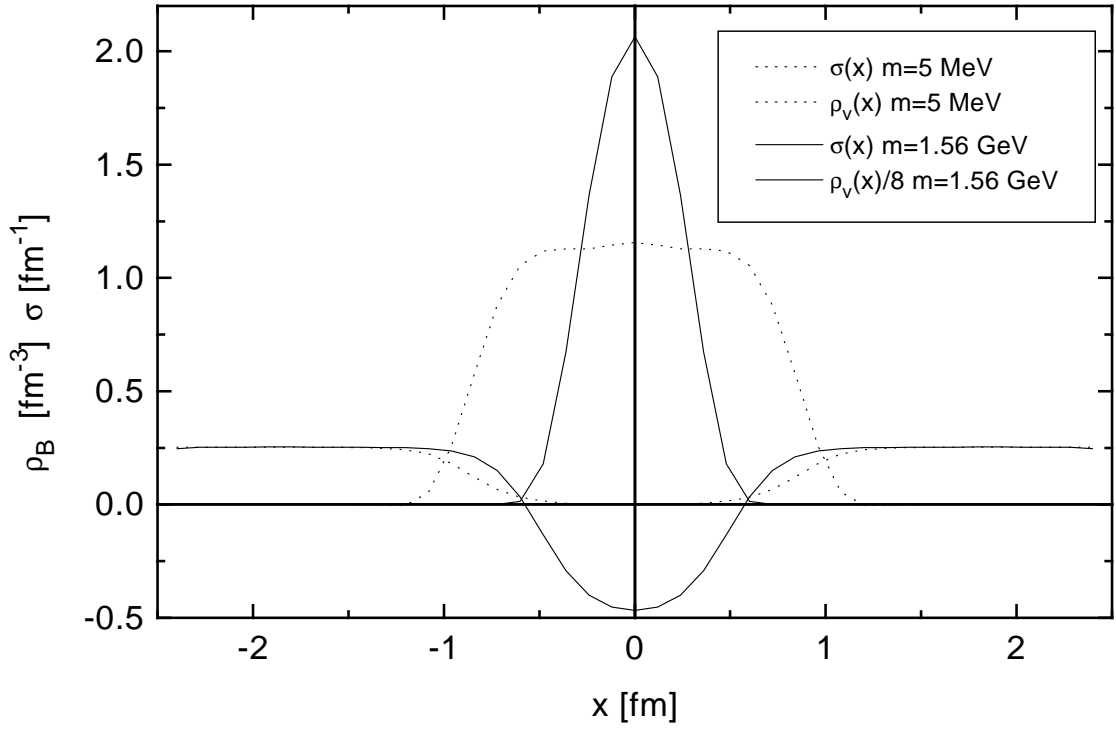,width=15cm}}
\vspace{2cm}
\caption{The radial distribution of the soliton field and the
quark density of the `heavy' quark of the $c\bar{c}$-state is depicted
at the initial time (when the constituent quarks are light) and
at a very late time (when the masses of the quarks have already incresed to their
final value).}
\label{figure2}
\end{figure}
\newpage
\begin{figure}
\vspace{0cm}
\centerline{\psfig{figure=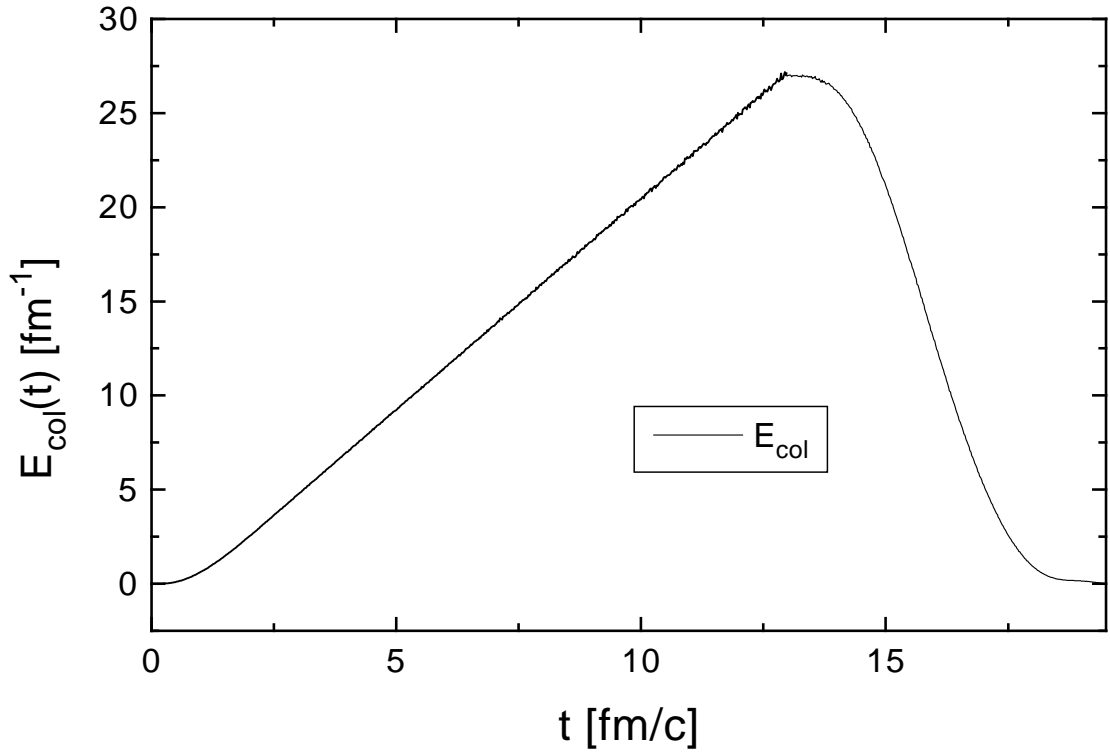,width=15cm}}
\vspace{2cm}
\caption{The temporal evolution of the total electric field energy inside
the string is shown. Up to times $t\approx 13$ fm/c the size of the
string is linearly increased to a final length of $\approx 8$ fm and
held constant afterwards.
At that time $t=13$ fm/c the $c\bar{c}$-state is inserted into the central region
of the string. A nearly linear decrease in time is observed.}
\label{figure3}
\end{figure}
\newpage
\begin{figure}
\vspace{-2cm}
\centerline{\psfig{figure=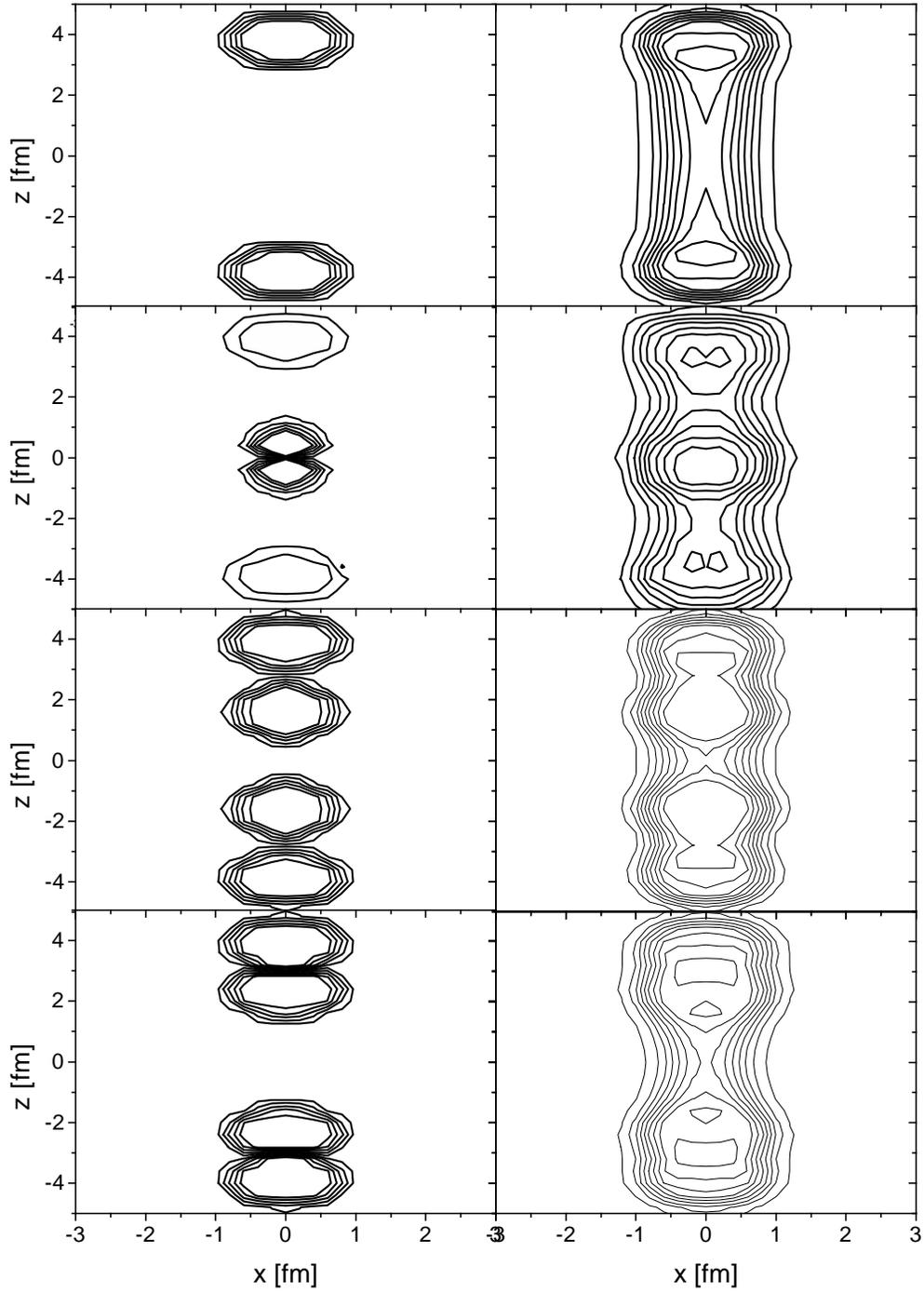,width=16cm}}
\vspace{-2cm}
\caption{\small The quark distribution (left side) and the scalar field
distribution (right side) at various times in the evolution
($t=12.8, \, 13.6, \, 16.2, \, 17.4$ fm/c) are shown in a contourplot.}
\label{figure4}
\end{figure}
\newpage
\begin{figure}
\vspace{-2cm}
\centerline{\psfig{figure=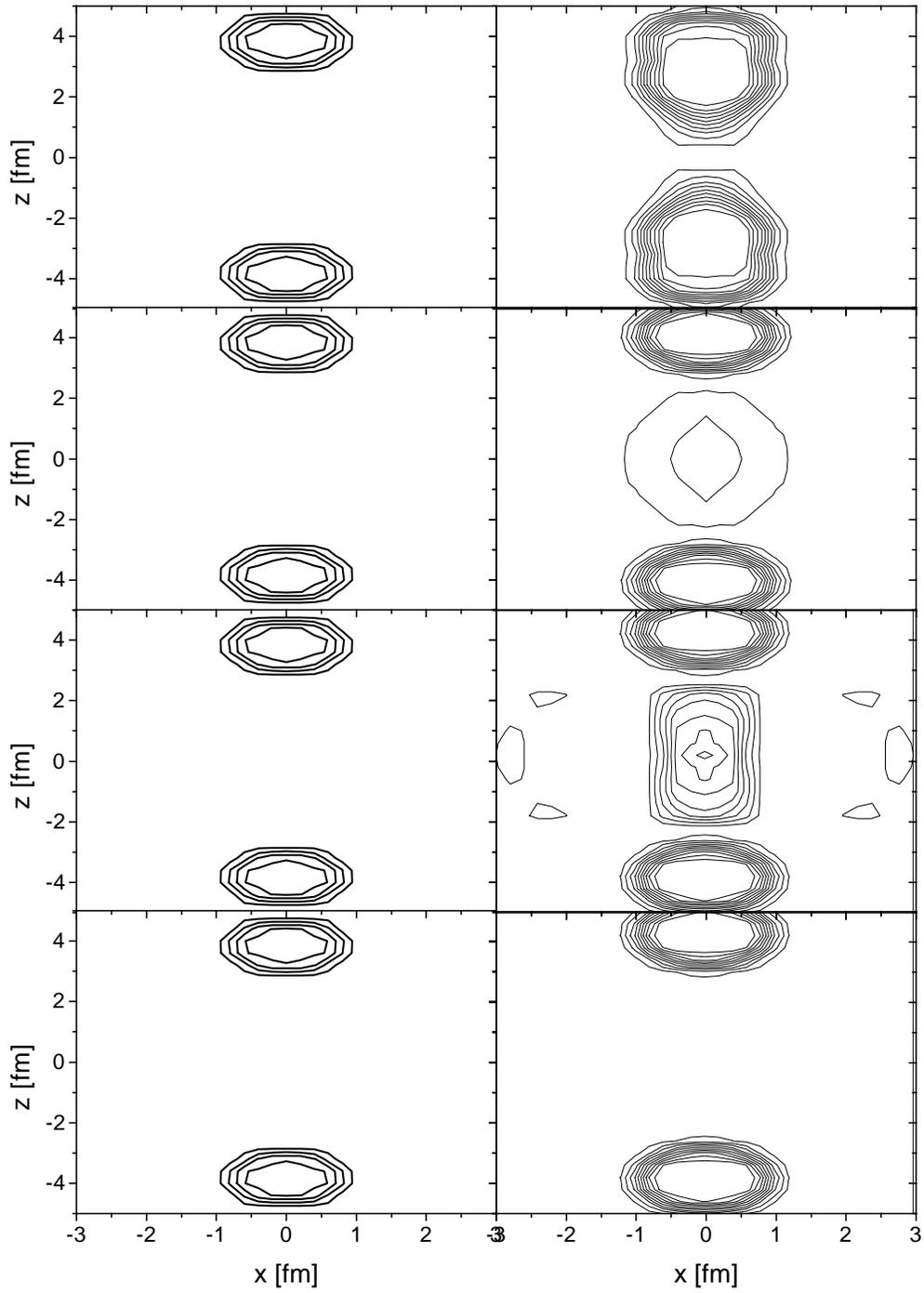,width=16cm}}
\vspace{-2cm}
\caption{\small The quark distribution (left side) and the scalar field
distribution (right side) at various times in the evolution
($t=18.6, \, 23.2, \, 26.8, \, 40.0$ fm/c) are shown in a contourplot.}
\label{figure5}
\end{figure}
\newpage
\begin{figure}
\centerline{\psfig{figure=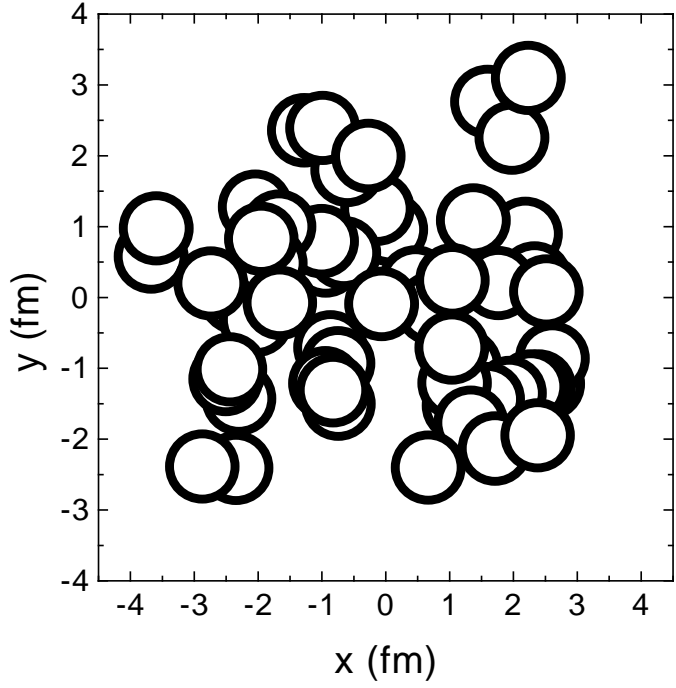,width=12cm}}
\caption{\small
The position of nucleonic strings is plotted,
all being produced in a cms-time intervall
of $\approx 0.5 $ fm/c within a slab of the complete transverse area
and 1 fm thickness along the longitudinal direction
in the cms frame in a central S+U reaction at 200AGeV generated
within the HSD algorithm \protect\cite{Ca96}. In addition, each string
is accompanied by a circle of radius
0.5 fm.}
\label{figure6}
\end{figure}
%
%
%
\end{document}